%% file: om.tex
\input harvmac
%\draftmode
\noblackbox

\input epsf
\input dnmacs

\def\tilde{\widetilde}
\newcount\figno
\figno=0
\def\fig#1#2#3{
\par\begingroup\parindent=0pt\leftskip=1cm\rightskip=1cm\parindent=0pt
\baselineskip=11pt
\global\advance\figno by 1
\midinsert
\epsfxsize=#3
\centerline{\epsfbox{#2}}
\vskip 12pt
{\bf Fig.\ \the\figno: } #1\par
\endinsert\endgroup\par
}
\def\figlabel#1{\xdef#1{\the\figno}}
\def\encadremath#1{\vbox{\hrule\hbox{\vrule\kern8pt\vbox{\kern8pt
\hbox{$\displaystyle #1$}\kern8pt}
\kern8pt\vrule}\hrule}}
\def\apm{{\alpha^{\prime}}}

\def\np#1#2#3{Nucl. Phys. {\bf B#1} (#2) #3}

%%% Paragraphs

%%% special math symbols
\font\cmss=cmss10
\font\cmsss=cmss10 at 7pt
\def\rlx{\relax\leavevmode}
\def\inbar{\vrule height1.5ex width.4pt depth0pt}
\def\IC{\relax\,\hbox{$\inbar\kern-.3em{\rm C}$}}
\def\IN{\relax{\rm I\kern-.18em N}}
\def\IP{\relax{\rm I\kern-.18em P}}
\def\ZZ{\rlx\leavevmode\ifmmode\mathchoice{\hbox{\cmss Z\kern-.4em Z}}
 {\hbox{\cmss Z\kern-.4em Z}}{\lower.9pt\hbox{\cmsss Z\kern-.36em Z}}
 {\lower1.2pt\hbox{\cmsss Z\kern-.36em Z}}\else{\cmss Z\kern-.4em
 Z}\fi}
%%% misc.
\def\IZ{\relax\ifmmode\mathchoice
{\hbox{\cmss Z\kern-.4em Z}}{\hbox{\cmss Z\kern-.4em Z}}
{\lower.9pt\hbox{\cmsss Z\kern-.4em Z}}
{\lower1.2pt\hbox{\cmsss Z\kern-.4em Z}}\else{\cmss Z\kern-.4em
Z}\fi}
%%% misc.
\def\IZ{\relax\ifmmode\mathchoice
{\hbox{\cmss Z\kern-.4em Z}}{\hbox{\cmss Z\kern-.4em Z}}
{\lower.9pt\hbox{\cmsss Z\kern-.4em Z}}
{\lower1.2pt\hbox{\cmsss Z\kern-.4em Z}}\else{\cmss Z\kern-.4em
Z}\fi}

\def\narrowplus{\kern -.04truein + \kern -.03truein}
\def\narrowminus{- \kern -.04truein}
\def\narrowminussub{\kern -.02truein - \kern -.01truein}

\def\half{{1\over 2}}

\def\cl{\centerline}

\def\b{{\beta}}

\def\a{{\alpha}}

\def\D{{\Delta}}
\def\m{{\mu}}
\def\n{{\nu}}
\def\ep{{\epsilon}}
\def\d{{\delta}}
\def\o{{\omega}}

\def\ph{{\phi}}
\def\t{{\theta}}
\def\T{{\Theta}}

\def\Ph{{\Phi}}

\def\r{{\rightarrow}}

\def\frac#1#2{{#1\over #2}}

\def\CL{{\cal L}}

\def\IZ{\relax\ifmmode\mathchoice
{\hbox{\cmss Z\kern-.4em Z}}{\hbox{\cmss Z\kern-.4em Z}}
{\lower.9pt\hbox{\cmsss Z\kern-.4em Z}}
{\lower1.2pt\hbox{\cmsss Z\kern-.4em Z}}\else{\cmss Z\kern-.4em
Z}\fi}
\def\IB{\relax{\rm I\kern-.18em B}}
\def\IC{{\relax\hbox{$\inbar\kern-.3em{\rm C}$}}}
\def\ID{\relax{\rm I\kern-.18em D}}
\def\IE{\relax{\rm I\kern-.18em E}}
\def\IF{\relax{\rm I\kern-.18em F}}
\def\IG{\relax\hbox{$\inbar\kern-.3em{\rm G}$}}
\def\IGa{\relax\hbox{${\rm I}\kern-.18em\Gamma$}}
\def\IH{\relax{\rm I\kern-.18em H}}
\def\II{\relax{\rm I\kern-.18em I}}
\def\IK{\relax{\rm I\kern-.18em K}}
\def\IP{\relax{\rm I\kern-.18em P}}
%\def\IX{\relax{\rm X\kern-.01em X}}
%this doesn't work

\def\p{\partial}

\font\cmss=cmss10 \font\cmsss=cmss10 at 7pt
\def\IR{\relax{\rm I\kern-.18em R}}

\def\me{M_{\rm eff}}
\def\ae{\apm_{\rm eff}}
\def\mp{M_{\rm p}}
\def\omt{{\dn\dnhalf :}~}
\def\aet{\tilde{\apm}_{\rm eff}}
\def\met{\tilde{M}_{\rm eff}}
%

%
%       \eqn\label{a+b=c}       gives displayed equation, numbered
%                               consecutively within sections.
%     \eqnn and \eqna define labels in advance (of eqalign?)
%
\def\eqnn#1{\xdef #1{(\secsym\the\meqno)}\writedef{#1\leftbracket#1}%
\global\advance\meqno by1\wrlabeL#1}
\def\eqna#1{\xdef #1##1{\hbox{$(\secsym\the\meqno##1)$}}
\writedef{#1\numbersign1\leftbracket#1{\numbersign1}}%
\global\advance\meqno by1\wrlabeL{#1$\{\}$}}
\def\eqn#1#2{\xdef #1{(\secsym\the\meqno)}\writedef{#1\leftbracket#1}%
\global\advance\meqno by1$$#2\eqno#1\eqlabeL#1$$}

%\lref\sstone{N. Seiberg, L. Susskind and N. Toumbas {\it The Teleological
%Behavior of Rigid Regge Rods}, hep-th/0005015. }
\lref\chu{\bibitem{Chu:1997iw}
C.~S.~Chu and E.~Sezgin,
``M-fivebrane from the open supermembrane,''
JHEP {\bf 9712}, 001 (1997)}
\lref\sas{S.~Kawamoto and N.~Sasakura,
``Open membranes in a constant 
C-field background and noncommutative  boundary strings,''
hep-th/0005123.}
\lref\li{M.~Li,
``Open membranes in matrix theory,''
Phys.\ Lett.\  {\bf B397}, 37 (1997)
hep-th/9612144.}
\lref\lst{N.~Seiberg,
``New theories in six dimensions and matrix description of 
M-theory on  T**5 and T**5/Z(2),''
Phys.\ Lett.\  {\bf B408}, 98 (1997)
hep-th/9705221.}
\lref\upan{Mandukya Upanishad, c.a. 700 B.C., see for instance {\it The
Principal Upanishads}, ed. S. Radhakrishnan, 1953 (Allen and Unwin).} 
\lref\sst{N.~Seiberg, L.~Susskind and N.~Toumbas,
``Strings in background electric field, space / time 
noncommutativity  and a new noncritical string theory,''
hep-th/0005040.}
\lref\bach{C. Bachas and M. Porrati, 
``Pair Creation of Open Strings in an Electric Field'',
Phys.Lett.{\bf B296}:77-84,1992.}
\lref\nest{V.V.Nesterenko, 
``The Dynamics of Open Strings in a Background Electromagnetic Field'', 
Int. J. Mod. Phys. {\bf A4} (1989) 2627-2652.}
\lref\callan{C.G. Callan, C. Lovelace, C.R. Nappi, S.A. Yost,
``Open Strings in Background gauge Fields'', 
Nucl.Phys.{ \bf B288} 525,1985. }
\lref\ft{E.S. Fradkin, A.A. Tseytlin,
``Nonlinear Electrodynamics from Quantized Strings'', 
Phys.Lett.{\bf B163} 123,1985.}
\lref\igor{I. Klebanov, Talk delivered at Lennyfest, Stanford, May 2000.}
\lref\burg{C. Burgess,
``Open String Instability in Background electric Fields'', 
Nuc. Phys. {\bf B294}427-444, 1987.}
\lref\cds{Alain Connes, Michael R. Douglas and Albert Schwarz,
``Noncommutative Geometry and Matrix Theory: Compactification on Tori'',
hep-th/9711162, JHEP 9802 (1998) 003.}
\lref\sw{N. Seiberg and E. Witten,
``String Theory and Noncommutative Geometry'', hep-th/9912072.}
\lref\gmms{R.~Gopakumar, J.~Maldacena, S.~Minwalla and A.~Strominger,
``S-duality and noncommutative gauge theory,''
hep-th/0005048.}
\lref\pole{E.~Cremmer and J.~Scherk, ``Factorization of the Pomeraon 
Sector, and currents in the dual resonace model,''
\np{50}{1972}{222}.} 
\lref\sav{O.~J.~Ganor, G.~Rajesh and S.~Sethi,
``Duality and non-commutative gauge theory,''
hep-th/0005046.}
\lref\asop{A.~Strominger,
``Open p-branes,''
Phys.\ Lett.\  {\bf B383}, 44 (1996)
hep-th/9512059.}
\lref\rwit{E.~Witten,
``Bound States Of Strings and p-Branes,''
Nucl.\ Phys.\  {\bf B460}, 335 (1996)
hep-th/9510135.}
\lref\gkp{S.~Gukov, I.~R.~Klebanov and A.~M.~Polyakov,
``Dynamics of (n,1) strings,''
Phys.\ Lett.\  {\bf B423}, 64 (1998)
hep-th/9711112.}
\lref\verl{H.~Verlinde,
``A matrix string interpretation of the large N loop equation,''
hep-th/9705029.}
\lref\wessez{P.S.~Howe, E.~Sezgin and P.C.~West, ``Covariant field
equations of the M-theory five-brane,'' Phys. Lett. {\bf B399} (1997)
49, hep-th/9702008; ``The Six-dimensional selfdual tensor,''
Phys. Lett. {\bf B400} (1997) 255, hep-th/9702111.}  
\lref\twonsend{P.~K.~Townsend, ``D-branes from M-branes,''
Phys.\ Lett.\  {\bf B373} (1996) 68, hep-th/9512062.}

\lref\chak{S.~Chakravarty, K.~Dasgupta, O.~J.~Ganor and G.~Rajesh,
``Pinned branes and new non Lorentz invariant theories,''
hep-th/0002175.}
\lref\bergs{E.~Bergshoeff, D.~S.~Berman, J.~P.~van der Schaar and
P.~Sundell, ``A noncommutative M-theory five-brane,''
hep-th/0005026.}
\lref\kasa{S.~Kawamoto and N.~Sasakura,
``Open membranes in a constant C-field background and noncommutative
boundary strings,'' hep-th/0005123.}%

\lref\sstc{N.~Seiberg, L.~Susskind and N.~Toumbas,
``Space/time non-commutativity and causality,''
hep-th/0005015.}
\lref\sethi{O.~J.~Ganor, G.~Rajesh and S.~Sethi,
``Duality and non-commutative gauge theory,''
hep-th/0005046.}
\lref\brn{J.~L.~Barbon and E.~Rabinovici,
``Stringy fuzziness as the custodian of time-space noncommutativity,''
hep-th/0005073.}
\lref\gome{J.~Gomis and  T.~Mehen, `` Space-Time Noncommutative Field 
Theories and Unitarity,'' hep-th/0005129.}
\lref\harmark{T.~Harmark,
``Supergravity and space-time non-commutative open string theory,''
hep-th/0006023.}
\lref\chenwu{G.~Chen and Y.~Wu,
``Comments on noncommutative open string theory: V-duality and  holography,''
hep-th/0006013.}
\lref\aloz{M.~Alishahiha, Y.~Oz and M.~M.~Sheikh-Jabbari,
``Supergravity and large N noncommutative field theories,''
JHEP {\bf 9911}, 007 (1999)
[hep-th/9909215].}
\lref\bersun{D.~S.~Berman and P.~Sundell,
``Flowing to a noncommutative (OM) five brane via its supergravity dual,''
hep-th/0007052.}
\lref\roylu{J.~X.~Lu, S.~Roy and H.~Singh,
``((F,D1),D3) bound state, S-duality and noncommutative 
open string /  Yang-Mills theory,''
hep-th/0006193.}
\lref\rusjab{J.~G.~Russo and M.~M.~Sheikh-Jabbari,
``On noncommutative open string theories,''
hep-th/0006202.}
\lref\juanig{I.~R.~Klebanov and J.~Maldacena,
%``1+1 dimensional NCOS and its U(N) gauge theory dual,''
hep-th/0006085.}
\lref\bergetal{E.~Bergshoeff, D.~S.~Berman, 
J.~P.~van der Schaar and P.~Sundell,
``Critical fields on the M5-brane and noncommutative open strings,''
hep-th/0006112.}
\lref\kawter{T.~Kawano and S.~Terashima,
``S-duality from OM-theory,''
hep-th/0006225.}
\lref\ahgom{O.~Aharony, J.~Gomis and T.~Mehen,
``On theories with light-like noncommutativity,''
hep-th/0006236.}

\lref\hashimoto{K.~ Hashimoto, ``Born-Infeld Dynamics in Uniform Electric 
Field,'' JHEP { \bf 9907} (1999) 016, hep-th/9905162, 
K.~Hashimoto, H.~Hata and S.~Moriyama,
``Brane configuration from monopole solution in non-commutative super
Yang-Mills theory,'' JHEP {\bf 9912} (1999) 021, hep-th/9910196;
A.~Hashimoto and K.~Hashimoto, ``Monopoles and dyons in
non-commutative geometry,'' JHEP {\bf 9911} (1999) 005,
hep-th/9909202;
K.~Hashimoto and T.~Hirayama,
``Branes and BPS configurations of noncommutative / commutative gauge
theories,'' hep-th/0002090.}
\lref\oth{L.~Jiang, ``Dirac monopole in non-commutative space,''
hep-th/0001073; D.~Bak,  
``Deformed Nahm Equation and a Noncommutative BPS monopole,''
hep-th/9910135, PLB 471 (1999) 149.}
\lref\sei{N.~Seiberg, ``Why is the Matrix Model Correct?,''
Phys.Rev.Lett.{ \bf 79}:3577-3580,1997, hep-th/9710009.}  

\Title{\vbox{\baselineskip12pt\hbox{hep-th/0006062}\hbox{IASSNS-HEP-00/48}
\hbox{}}}{ {\dn\dnhuge :}~(OM)  Theory in Diverse Dimensions}

 \centerline{ Rajesh ${\rm Gopakumar}^1$, Shiraz ${\rm Minwalla}^1$, 
Nathan ${\rm Seiberg}^2$ and Andrew ${\rm Strominger}^1$}
\bigskip\centerline{$^1$ Jefferson Physical Laboratory}
\centerline{Harvard University, Cambridge, MA 02138}
\smallskip
\centerline{$^2$ School of Natural Sciences} 
\centerline{Institute for Advanced Study,
Princeton, NJ 08540} 

\vskip .3in \centerline{\bf Abstract} {
Open string theories can be decoupled from
closed strings and gravity by scaling to the critical electric field.
We  propose dual descriptions for the strong coupling limit of these NCOS 
(Non-Commutative Open String) theories in  six or fewer spacetime dimensions.
In particular, we conjecture that the five-dimensionsal NCOS theory 
at strong coupling, is a theory of light Open Membranes (OM={\dn\dnhalf :}), 
decoupled from gravity, on an M5-brane with a near-critical 
three-form field strength. The relation of \omt theory to NCOS theories
resembles that of M theory to Type II 
closed string theories. 
In two dimensions 
we conjecture that supersymmetric U($n$) gauge theory 
with a unit of electric flux is 
dual to the NCOS theory with string coupling $1 \over n$. A construction 
based on NS5-branes leads to new
theories in six dimensions generalising the little string theory. A web of 
dualities relates all the above theories when they are compactified on tori.}

\smallskip
\Date{} 
\listtoc 
\writetoc
\newsec{Introduction}

It has been a long-held belief that open string theories always
require closed strings for consistency at the quantum level, due
to the appearance of poles in one-loop open string scattering
amplitudes \pole. This belief has recently been questioned.
Weakly-coupled theories of open strings on D-branes 
were constructed by scaling to a critical electric 
field, and S-duality was used to argue that they
decouple from closed strings\refs{\sst, \gmms}. The decoupling 
was verified, for two through six dimensional branes
(IR problems may appear for higher dimensions), by 
the absence of closed string poles in nonplanar loop diagrams \gmms .
These simplified string theories thus permit the investigation
of mysterious stringy phenomena without the
complications of gravity and consequent loss of a fixed background
geometry. As the name NCOS theory (Non-Commutative Open
String) indicates, they exhibit non-commutativity of space and time 
coordinates (spacetime noncommutativity was also considered in
\refs{\sstc, \sav, \brn, \gome, \chenwu } . The corresponding supergravity
solutions are studied in  \refs{\aloz , \harmark }).
In five and six dimensions they also provide a
non-gravitational ultraviolet completion of Yang-Mills theory.

We expect that these theories are part of a web of theories
related by duality and compactification. In this paper we explore
a piece of this web by seeking strong coupling duals for all the
NCOS theories. In four dimensions it was already argued in \gmms\
(see also \sav) that the NCOS theory is dual to spatially
noncommutative, maximally supersymmetric
 Yang-Mills field theory. In five dimensions we conjecture that the strongly
coupled NCOS theory consists of M5-branes with a near-critical
three-form field 
strength.

The M5-brane is the boundary of fluctuating open
membranes, much as 
D-branes are the boundaries of fluctuating open strings 
\refs{ \asop, \twonsend}.  
Near criticality these open
membranes become nearly tensionless.
This theory --
{\dn\dnhalf :} (OM\foot{{\dn :}~(OM): That which captures 
the underlying nature of
reality \upan, or Open Membrane,  according to taste.})
 theory -- is described
by the gravitationally  decoupled
dynamics of the light open membranes \foot{Just as for M-theory, 
we will interchangeably use the term \omt
theory also for the whole web of non-gravitational theories related via
compactifications and dualities to the theory with light open membranes.}.
 
The M5-brane near a critical three form field, and its 
compactifications were considered in \sw, however we do not 
consider the rank four case of that paper.
Decoupled theories with constant
$C$ have also been explored by various authors including
\refs{\sw,\aloz, \chak,\bergs,\kasa}.  It has 
 been conjectured that just
as nonzero $B$ is related to noncommutativity, nonzero $C$ might be
related to nonassociativity.  However, it is not clear how to make
this conjecture precise.

We go on to define a large class of new six 
dimensional non-gravitational theories with light open D-branes among their 
excitations. Specifically, these are scaling theories on NS5-branes with near
critical RR gauge fields of different ranks. This results in the presence
of corresponding light branes in the spectrum.
These are part of the \omt web of theories, being related by 
various dualities on circle compactification.

The
two-dimensional NCOS theory (see \refs{\verl, \gkp,\igor}) has the unique
feature that the open string coupling is quantized and 
bounded, $G_o^2 ={1 \over
n} \leq 1$; thus there is no strong coupling limit. However we argue
that the two dimensional NCOS theory at weak coupling (large $n$) 
 is dual to strongly coupled two-dimensional U($n$) gauge theory with 
discrete electric flux. We argue that the strong coupling 
limit of the three dimensional NCOS theory is the SO(8) invariant  
M2-brane worldvolume field  theory.  

{\it Note added to revised edition}: Related independent work has  
appeared in \bergetal . Several papers analysing the properties of 
\omt theory have also appeared since the original version. 

\newsec{OM Theory} 

In this section we will consider the theory of an M5-brane in the presence
of a near critical electric $H_{012}$ field. We will find that  
in the limit ${H_{crit} -H \over H_{crit}} \to 0$, the tension of open 
membranes stretched spatially in the $1,2$ directions is infinitely 
below the Planck scale. It is thus possible 
to define a theory of light fluctuating 
open membranes propagating on the M5-brane, decoupled
from gravity. Like the (0,2) theory, \omt theory 
has no dimensionless parameters, 
and so is unique and strongly coupled. In fact \omt theory 
reduces to the (0,2) field theory at low energies.

Consider M theory 
in the presence of N coincident M5-branes with  
a background worldvolume 3-form field strength  
\eqn\trh{H_{012}=\mp^3\tanh \beta,}
and an asymptotic metric 
\eqn\grt{g_{\m\n}=\eta_{\m\n},~~~~~g_{ij}=f^2\delta_{ij},~~~~~
 g_{MN}=h^2 \d_{MN} ,}
with $\m,\n=0,1,2~~~i,j=3,4,5, ~~~ M, N=6, 7, 8, 9, 11$. 
$f$ and $h$ are constants introduced for later convenience. 
The nonlinear self duality constraints \wessez\ then determine the other 
components of $H$ as 
\eqn\trdh{H_{345}=-f^3\mp^3 \sinh \beta.}
The effective tension of a membrane (proper 
mass per unit proper area) stretched spatially in the 1,2 directions is 
\eqn\tyj{{1\over
4\pi^2}\left(\mp^3-\ep^{012}H_{012}\right)
={\mp^3 \over 4\pi^2 e^{\b} 
\cosh \b  }\equiv {1\over 4\pi^2}\me^3.}
$\mp$ is the gravitational scale while $\me$ 
sets the scale for the proper energies of 
fluctuations of these open membranes. 
As $H_{012}$ is scaled to its critical value (i.e $\b$ is taken to 
$\infty$), ${ \mp \over \me}\sim e^{{2 \b\over 3}} \r \infty$, 
and the fluctuating open membranes decouple from gravity!
In order to focus on these light modes we take 
the limit as $\mp \sim e^{{2\beta\over 3}} \r \infty, $
$\me$  fixed.

%The energy of membranes stretched in the 1-2 spatial 
%directions is given by 
%\eqn\ten{p_0=\me^3 \ep_{012}\D x^1 \D x^2.}

We have judiciously chosen $\m,\n$ coordinates in \grt\  so that the 
energy per unit coordinate area of a membrane aligned along the 012 
direction is finite. This condition does not fix the $i,j$ coordinate
system or equivalently a 
choice of $f$ in \grt. We fix this by demanding that a membrane
along $e.g.$ 034 has finite energy per unit coordinate area. Since the 
energy per 
unit proper area of such a membrane 
(as measured by $g_{ij}$) diverges like $\mp^3$ in the 
limit, this requires a small $f$: 
\eqn\findr{ \ep_{034}\mp^3={\rm finite} \Rightarrow 
f^2\propto{\me^3 \over \mp^3 }.}
For later convenience we make the specific choice\foot{It is possible that
the factor of 2 can be motivated from isotropy of the ``open membrane
metric''\bergs\ but we shall not do so here.}
\eqn\rfc{f^2=2{\me^3 \over \mp^3}.} 
It is also convenient to choose 
\eqn\tcords{h^2=f^2=2{\me^3 \over \mp^3}.}
Below we will argue that, with this choice of transverse coordinates, 
the dimension two scalar operators $\Ph^M$ (normalized to have the 
usual kinetic term) representing the transverse 
fluctuations of the 5-brane in the low energy field theory limit of 
OM theory, are related to the geomertrical position of the 5-brane
by a factor of $\me^3$; $\Ph^M \sim \me^3 X^M$. 

In summary, we consider $N$ M5-branes in the \omt limit
\medskip
\cl{Table 1}
\bigskip
\centerline{\vbox{\offinterlineskip
\hrule
\halign{&\vrule#&
        \strut\quad#\quad\cr
height3pt&\omit&&\omit&\cr &The
\omt Limit \hfil & \cr height3pt&\omit&&\omit&\cr \noalign{\hrule}
height3pt&\omit&&\omit&\cr & $\mp^3 ={\me^3 \over 2}e^{2\beta}$ \hfil &\cr
&  $H_{012}=\mp^3 \tanh \beta$ \hfil &\cr
& $g_{\m\n}=\eta_{\m\n}$ \hfil & \cr
& $ g_{ij}= { 2\me^3 \over \mp^3} \delta_{ij}$ \hfil & \cr
& $g_{MN} = { 2\me^3 \over \mp^3} \delta_{MN}$ \hfil & \cr
height3pt&\omit&&\omit&\cr}
\hrule} }
$$\b \r \infty ~~~~~~ \me ~~{ \rm fixed}, ~~(\m,\n=0,1,2, ~~~
i,j=3,4,5 , ~~~M,N=6,7,8,9,11.)$$
\medskip 
%\eqn\scalinglimit{\eqalign{\mp^3 =&{\me^3 \over 2}e^{2\beta}, ~~~
%H_{012}=\mp^3 \tanh \beta, ~~~
%g_{\m\n}=\eta_{\m\n},~~~g_{ij}= { 2\me^3 \over \mp^3} \delta_{ij}, 
%~~~g_{MN} = { 2\me^3 \over \mp^3} \delta_{MN}, \cr 
%& \b \r \infty ~~~~~~ \me ~~{ \rm fixed}, ~~(\m,\n=0,1,2, ~~~
%i,j=3,4,5 , ~~~M,N=6,7,8,9,11.) \cr}}
The resultant \omt theory contains fluctuating open membranes of proper 
energy $\CO(\me)$, decoupled from gravity. Note that \omt theory has 
no dimensionless parameters. 
We will argue below that \omt theory, upon 
compactification in the 2 direction, reduces to the 4+1 dimensional NCOS 
theory, and therefore is not a trivial theory, even in the case $N=1$.
Thus, for simplicity, 
we will concentrate on the theory of a single 5-brane through the 
rest of this paper, although our considerations may 
be generalized.

\newsec{Review of the NCOS Limit}

In subsequent sections we will study OM theory compactified on 
various circles. In particular, we will 
find a relationship between OM theory and the 4+1
dimensional NCOS theory. 
In this section we review the NCOS limit in coordinates 
convenient for present purposes.\foot{Our conventions and coordinates
here
 differ from those 
employed in \gmms, which were chosen to elucidate the relation to the 
$S$-dual field theory.}
We will also discuss the T-duality of NCOS theories. 
   
Consider a Dp-brane with a near critical electric field
in the $0,1 \equiv \m,\n$ direction and closed string coupling 
denoted by $g_{str}$. The closed string metric and electric field  
can be chosen as 
\eqn\csq{g_{\m\n}= \eta_{\m\n},~~~~g_{ij} = \ep  \d_{ij},~~~~
g_{MN}=\ep \d_{MN}, 
~~~2 \pi \apm\ep^{01}F_{01}=1-{\ep\over 2},~~~}
with $\ep \ll 1$ , $i,j=2,3, \cdots , p$ the non-electric  directions on the 
brane, and $M,N=p+1, \cdots , 9$ the directions transverse 
to the brane. 
Using the formulae in \refs{\ft,\callan,\sw} it is easy to determine the 
open string metric,
noncommutativity parameter and string coupling $G_o^2$
corresponding to the closed string moduli of \csq,
\eqn\osq{G_{\m\n}={\ep}\eta_{\m\n}, ~~~G_{ij}=\ep \d_{ij}, ~~~ \T^{\m\n}
={2\pi\apm \over \ep }\epsilon^{\m\n} ,~~~ G_o^2=g_{str}\sqrt{{\ep}} .}

The effective tension (energy per unit length)
of a string stretched in the 1 direction is  
\eqn\stringten{{1 \over 2 \pi}
\left({1\over \apm}-2\pi \ep^{01} F_{01}\right)={\ep \over 4\pi \apm} 
 \equiv {1\over 4 \pi \ae}.}
$\apm$ sets the scale of closed string oscillators, 
and $\ae$ the scale for the  
energy of oscillating open strings. As the electric field is scaled to 
its critical value, ${\apm \over \ae} = {\ep} \r 0$, and the oscillating 
open strings decouple from gravity. We take $\ae$ fixed as $\ep$ $\r 0$, 
so that $\apm \propto \ep  \r 0$.
Open string oscillator states obey the mass shell condition 
\eqn\osms{p_{A} \eta^{AB} p_{B} ={N \over \ae}}
and so have proper energy 
$p_0^2= \CO({1\over \ae})$ as expected.
The part of the string sigma model involving transverse coordinates 
is  
\eqn\ssmdl{ S= { 1\over 4 \pi \apm} \int g_{MN} \p X^M \bar{\p} X^N
={ 1 \over 4 \pi \ae} \int \d_{MN} \p X^M \bar{\p} X^N.}
Thus correlation functions of the $X^M$ fields are finite as $\apm \r 0$, 
and the dimension one scalar fields $\ph^M$ (normalized to have
standard kinetic term) in the low energy gauge theory 
on the NCOS brane worldvolume, are related to the coordinates 
$X^M$ by a factor of 
${1 \over \ae}$; $\ph^M \sim  {X^M \over \ae}$.  
Finally, we scale $g_{str}$ to keep 
$G_o^2=g_{str} \sqrt{{ \apm \over \ae}}$ fixed as $\apm$ is taken to 
zero. This limit (the NCOS limit) results in a one parameter 
family of  interacting  open string theories, 
(NCOS theories), labelled by their coupling constant 
$G_o$ and decoupled from gravity.
\medskip
\cl{Table 2}
\bigskip
\centerline{\vbox{\offinterlineskip
\hrule
\halign{&\vrule#&
        \strut\quad#\quad\cr
height3pt&\omit&&\omit&\cr &The
NCOS Limit \hfil & \cr height3pt&\omit&&\omit&\cr \noalign{\hrule}
height3pt&\omit&&\omit&\cr & $g_{\m\n}=\eta_{\m\n}$ \hfil &\cr
&  $g_{ij }={\apm \over \ae}\d_{ij} $\hfil &\cr
&  $g_{MN }={\apm \over \ae}\d_{MN} $\hfil &\cr
& $2\pi\ep^{01}\apm F_{01}=1-{\apm \over 2\ae}$ \hfil & \cr
& $g_{str}=G_o^2\sqrt{{\ae \over \apm}}$ \hfil & \cr
& ${G}^{AB}={\ae \over \apm}\eta^{AB}$ \hfil & \cr
& $\T^{\m\n}=2\pi\ae \ep^{\m\n}$ \hfil & \cr
height3pt&\omit&&\omit&\cr}
\hrule} }
\medskip  

At low energies, the NCOS theory reduces to 
\eqn\len{\eqalign{S&={1 \over 4 (2\pi)^{p-2}G_o^2 \apm^{{p-3 \over 2}}} 
\int d^{p+1} x \sqrt{-G}G^{AM}G^{BN}\hat{F}_{AB}\hat{F}_{MN} \cr
&={ 1 \over 4 (2 \pi)^{p-2} G_o^2 \ae^{{p-3  \over 2}}}
\int d^5 x  {\eta}^{AM} {\eta }^{BN}\hat{F}_{AB}\hat{F}_{MN} \cr}}
i.e. it reduces to Yang Mills theory with 
$g^2_{YM}= (2 \pi)^{p-2} G_o^2 
\ae^{{p-3  \over 2}}$.

\subsec{T Duals of NCOS Theories}

In this section we review the action of T-duality on the NCOS theories for
later use \foot{This discussion has independently appeared in \kawter\ .}.
Consider a $p+1$ dimensional brane in the NCOS limit of 
Table 2,  wrapped on a circle of 
coordinate radius $R$ in the $p^{th}$  spatial direction. 
Performing a T-duality in the $p^{th}$ direction yields a 
$(p-1)+1$ dimensional brane, at a point on the now transverse 
$p^{th}$ circle, whose coordinate radius is given by 
\eqn\clatd{\tilde{R}= { \ae \over \apm} \times {\apm \over R}=
{\ae \over R}.}
The asymptotic value of the string coupling after T-duality, 
$g'_{str}$,  is given by
\eqn\avsc{g'_{str}=g_{str}{\sqrt{\apm} \over  \sqrt{\apm \over \ae} R}
=g_{str}{ \sqrt{\ae} \over R} ={G_o^2 \sqrt{\ae} \over R}
\sqrt{ {\ae \over \apm}}.}
The asymptotic values of the metric, and the worldvolume electric field 
are unchanged by the T-duality, and so may be read off from Table 2. 

In conclusion, the $p+1$ dimensional NCOS theory with scale $\ae$ and 
coupling $G_o$, wrapped on a circle of coordinate length $R$ in 
a non-electric direction, is T-dual to a $(p-1)+1$ dimensional 
NCOS theory with one compact transverse scalar.
The $(p-1)+1$ dimensional NCOS theory has scale $\ae$ and coupling $G_o^{'2}= 
{G_o^2 \sqrt{\ae} \over R}$. The coordinate radius of 
compactification of the transverse scalar is $\ae \over R$
\foot{ The dimension one field $\ph^p$ (normalized to have the usual kinetic
term) in the low energy gauge theory is compact with radius 
$ \sim {1 \over R}$. }.
Notice that the NCOS radius and coupling transform under T-duality  
exactly as the analogeous closed string quantities 
transform under the usual closed string T-duality, with $\ae$ playing
the role of $\apm$.

\newsec{Compactification of OM theory on an Electric Circle} 

Consider \omt theory compactified on a spatial circle 
of proper (and coordinate) radius $R$ 
in one of the `electric' spatial directions (the direction $x^2$
for definiteness). Since 
\omt theory reduces to the (0,2) theory at energies 
well below $ \me$, 
the low energy dynamics of the compactified theory is governed by 
4+1 dimensional Yang Mills with $g_{YM}^2 \sim  R$. 
At higher energies light
open membranes wrap the compactification circle to form 
the light open strings decoupled from gravity. 
It is natural to guess that this theory is the 5 dimensional  
NCOS theory with effective string tension
${1\over 4\pi\ae} \sim \me^3 R$ 
and open string coupling $G^2_{o}\sim {g^2_{YM}\over 
\sqrt{\ae}}
\sim (R \me)^{3\over 2}$. In this section we will verify
that this is indeed  the case. 

\omt theory compactified on a spatial circle (say in the 2 direction)
of proper radius $R$, may be obtained as follows. Consider
M theory on $S^1 \times R^{10}$ (the $S^1$ is in the 2 direction)  
with M5-branes wrapping the circle. Scale all bulk moduli 
as in Section 2;
in particular  
\eqn\au{\eqalign{&\ep^{012}H_{012}=\mp^3 -\me^3,~~~ g_{\m\n}=\eta_{\m\n}  
~~(\m,\n=0,1,2),~~g_{ij}={2\me^3 \over \mp^3}\d_{ij}~~ (i,j=3,4,5),\cr
& ~~~~~~~ g_{MN}={2\me^3 \over \mp^3}\d_{MN}~~ (M,N={\rm transverse}),
~~~\mp \r \infty,  ~~~\me ~~{ \rm fixed.} \cr}}
The dictionary between M-theory and IIA implies that this system
is equivalent to a D4-branes in IIA theory with
\eqn\auu{\eqalign{&{1 \over {\apm}} ={R\mp^3}, ~~
g_{str}=(R\mp)^{{3 \over 2}}, ~~g_{\m\n}=\eta_{\m\n}~~(\m,\n=0,1),~~~ 
g_{ij}= {2\me^3 \over \mp^3 }\d_{ij}~~(i,j=3,4,5),\cr
&~~~g_{MN}= {2\me^3 \over \mp^3 }\d_{MN}~~(M,N={\rm transverse}),~~~
2 \pi \apm F_{01}=\apm {R} H_{012}=1-{\me^3 \over  \mp^3}.}}
In the limit $\mp \to \infty$, comparing with Table 2, 
 we find ourselves in the NCOS limit,  
$\apm \to 0$  with  
fixed effective open string tension and open string coupling   
\eqn\av{{ 1 \over 2\ae}={\me^3 \over \mp^3\apm} =\me^3 R, ~~~G_o^2=
g_{str}\sqrt{{2\me^3 \over \mp^3}}=\sqrt{2}(R\me)^{{3\over 2}}.}

Thus the 4+1 dimensional NCOS theory with 
scale $\ae$ and coupling $G_o^2$ may be identified with 
\omt theory with scale $\me$, compactified on an electric circle of 
radius $R$  with 
\eqn\scf{ \me={2^{-{1 \over 3}} \over \sqrt{\ae}G_o^{{2 \over 3}}}, ~~~
R={G_o^2 \sqrt{\ae}.}}

The relation between \omt theory and the 5 dimensional 
NCOS theory is reminiscent of the relationship between 
M theory and IIA string theory. Notice that the radius of the 
compactification circle 
in units of $1 / \me$ is equal to $G_o^{{4 \over 3}}$.
Thus, at strong NCOS coupling  Kalutza Klein modes are much lighter 
than $\me$ and  
the theory is effectively 6 dimensional. 

As argued in the previous section, the dimension one scalar field $\ph^M$ 
(normalized to have unit kinetic term) on the worldvolume of the 
NCOS brane is related to transverse coordinate position, 
by $\ph^M \sim { X^M \over \ae}$. 
However, the corresponding dimension 2 field $\Ph^M$ 
on the OM worldvolume, is related to $\ph^M$ by $\Ph^M \sim {\ph^M \over
R}$.
Combining these formulae, and using \scf\ we conclude that 
$\Ph^M \sim \me^3 X^M$, as asserted in section 2.

\newsec{Compactification of OM theory on a Magnetic Circle}

Again consider \omt theory compactified on a spatial circle,  
this time of { \it coordinate} radius $L$  (proper radius 
$\sqrt{2 \me^3 \over \mp^3} L$)
in one of the `magnetic' spatial directions (the direction $x^3$
for definiteness). As in the previous subsection,
the compactified theory at low energies is 
$4+1$ dimensional SYM with gauge coupling $g_{YM}^2\sim L$. 
Indeed, we will see below that 
the effective $4+1$ dimensional description 
of this theory is 4+1 dimensional noncommutative SYM.
Since noncommutative SYM  is nonrenormalizable in five dimensions,
the theory does not have a complete 4+1 dimensional description.
\omt theory on a circle provides a
completion of 4+1 dimensional noncommutative SYM.

Proceeding as in the previous section, we find that the compactified 
theory may equivalently be described as a D4-brane 
in IIA theory with parameters
\eqn\aww{\eqalign{\apm=&{ 1 \over L \sqrt{ 2 \me^3 \mp^3 }}~~~
g_{str}= \bigl({ 2L^2 \me^3 \over \mp})^{{3 \over 4}},~~~
g_{\m\n}=\eta_{\m\n} ~~(\m,\n=0,1,2), \cr
&g_{ij}=2 {\me^3 \over \mp^3}\d_{ij}~~(i,j=4,5), 
~~~ F_{45}={ L\me^3 \over \pi}.\cr}}
As in \sw, (the parameter $\ep$ of \sw\ may be identified 
with $e^{-2\b}$ in \aww)
the decoupled field theory on the D4-brane is maximally 
supersymmetric  U(1) noncommutative 
Yang Mills (NCYM) with open string metric
\eqn\mom{G_{\m\n}= \eta_{\m\n}, ~~~
G_{ij}={(2 \pi \apm F_{45})^2  }{\mp^3 \over 2 \me^3}\d_{ij}= \d_{ij}}
and noncommutativity 
\eqn\mnc
{\T^{ij}={ \epsilon^{ij} \over F_{45}}={ \pi \epsilon^{ij} \over L \me^3}.}
At low energies, the NCYM theory is governed by the Lagrangian
\eqn\lag{\CL= {1 \over 4g_{YM}^2}\int d^5 x 
\sqrt{-G}
 G^{AM}G^{BN}\hat{F}_{AB}\hat{F}_{MN}  }
where
\eqn\mym{g^2_{YM}=(2 \pi)^2 { \sqrt{\apm}g_{str} }\sqrt{   \det G_{ij}\over 
\det  (2 \pi \apm F_{ij})} = 4 \pi^2 L} 
as expected.

\subsec{Compactification on an Electric and a Magnetic Circle}

Another interesting compactification which combines the two previous
ones is on the circle $(x^2,x^5) \sim (x^2,x^5) + (2\pi L_2,2\pi L_5) $. (The
following discussion has also appeared
independently in \rusjab. See also \refs{\chenwu, \roylu , \kawter}). 
Since
in the metric $g$ the distances along $x^5$ are scaled to zero, the
radius of the circle in the scaling limit is independent of $L_5$.  It
is given by $L_2$, and therefore $\alpha'$ and the string coupling are
as in the compactification on the electric circle.  The two form in
the noncompact directions is determined as $F=\oint
H=2\pi L_2H_{012}dx^0dx^1+ 2\pi L_5H_{345}dx^3dx^4$.

We start by analyzing the situation in directions $0,1$.  Since the
electric field scales to the critical value as in the compactification
on the electric circle, the open string metric $G$ and the
noncommutativity parameter $\Theta$ in these directions are as in that
problem, i.e.\ $G$ scales like $1\over M_p^3$ and $\Theta^{01}$ is
finite.

In directions $3,4$ the situation differs from that in the previous
cases.  Here $g$ scales like $1\over M_p^3$ and $F$ is of order one.
Since $\alpha'$ is of order $1\over M_p^3$, the two terms in the
denominator of $G^{-1} + {1 \over 2\pi \alpha'} \Theta= {1 \over g+
2\pi \alpha' F}$ are of the same order of magnitude.  We conclude that
the components of $G$ in these directions are of order $1\over M_p^3$
and that $\Theta^{34}$ is of order one.

Finally, there is one more noncompact direction.  It is
straightforward to check that the induced metric along that direction
is of order $1\over M_p^3$, and therefore $G$ is also of that order.

We conclude that all the components of $G$ are of order $1\over M_p^3$
and that $\Theta^{01}$ and $\Theta^{34}$ are of order one.  This is
similar to the situation with the electric circle except that there is
also noncommutativity in the spatial directions; i.e.\ $\Theta$ is of
rank four.

\newsec{Near Critical NS5-brane Theories}

In this section we will define a series of new six dimensional
theories, ODp (Open Dp-brane) 
theories\foot{ We have learnt of independent related work by 
T. Harmark (to appear).}
, labelled by an integer $p$ where $p$ runs
from 1-5.  These are decoupled theories on the world volume of the
NS5-brane; the excitations of these theories include light open
Dp-branes.  These Dp-branes remain light, even in the decoupling
limit, due to the presence of a near critical p+1 form Ramond-Ramond
potential (NS5-branes with background RR fields were studied by
various authors including
\ref\cederwall{M.~Cederwall, U.~Gran, M.~Nielsen and B.~E.~Nilsson,
``(p,q) 5-branes in non-zero B-field,''
JHEP {\bf 0001} (2000) 037, [hep-th/9912106].}).

As is well known, all even Dp-branes can end on NS5-branes in the IIA
theory and all odd Dp-branes can end on NS5-branes in the IIB theory
(when $p < 6$).
Therefore, with an appropriate background RR field such open Dp-branes
can be made very light.  This is similar to the light open fundamental
strings on D-branes in the NCOS theories and the light open membranes
of \omt theory, discussed in the previous sections.  

Our construction of near critical NS5-brane theories 
can be generalized by turning on several different
Ramond-Ramond fields.  In these theories the 
light D-branes carry several charges.
We will not analyze these generalized theories in detail in this paper.

The various near critical NS5-brane theories are distinct 
six dimensional theories but, upon compactification, all lie on the 
same moduli space. This is similar to the
equivalence of the type IIA and type IIB string theories when
compactified on a circle and to the equivalence of the IIA and IIB
little string theories \lst.
As we will argue below, ODp theories are also on the same moduli space 
as NCOS theories and \omt theory.  

In order to motivate our construction, consider a D5-brane in IIB
theory, in the $5+1$ dimensional NCOS limit, as given in Table
2. S-dualizing this background yields a scaling limit that defines the
OD1 theory, a decoupled theory on the world volume of the
IIB-NS5-brane, whose excitations are open D-strings. These are light
in the decoupling limit because of a near critical background 
$C_{01}$ RR potential. Compactifying this theory on tori,
T-dualizing, and decompactifying the resultant theories, yields the
scaling limits that define the various ODp theories.  These scaling
limits are summarized in Table 3 below.
\medskip
\cl{Table 3}
\bigskip
\centerline{\vbox{\offinterlineskip
\hrule
\halign{&\vrule#&
        \strut\quad#\quad\cr
height3pt&\omit&&\omit&\cr &~~~~~~
ODp Theories \hfil & \cr height3pt&\omit&&\omit&\cr
\noalign{\hrule} 
height3pt&\omit&&\omit&\cr & $g_{\m\n}=\eta_{\m\n}, ~~\m,\n
=0,1,\ldots p$  \hfil &\cr
&  $g_{MN }=\epsilon\delta_{MN}, ~~~M,N =(p+1),\ldots 9$\hfil &\cr
& $\tilde{\apm}=\epsilon^{1\over 2}\aet$ \hfil &\cr
& $g_{str}^{(p)}=\epsilon^{3-p\over 4}G_{o(p)}^2$  \hfil & \cr
& $\ep^{01\ldots p}C_{01\ldots p}={1\over (2\pi)^p
G_{o(p)}^2{\aet}^{p+1\over 2}}({1\over \epsilon} -{1\over 2})$ \hfil
&\cr  
& $C_{(p+1)\ldots 5}={1\over (2\pi)^{4-p}G_{o(p)}^2{\aet}^{5-p\over
2}}$ \hfil & \cr
height3pt&\omit&&\omit&\cr}
\hrule} }
\medskip 
In the table above $\tilde{\apm}$ and  $g_{str}^{(p)}$ represent 
the closed string scale and closed string coupling respectively. 
Note that $M, N$ run over all dimensions transverse to the brane,
as well as the brane directions orthogonal to the critical 
$C$ field.

Like the NCOS theories, the ODp theories are labelled by two
parameters, the dimensionless $G_{o(p)}$ and a scale $\aet$. Below we
will comment on the interpretation of these parameters.

As an aside we note that in  Table 3 we have made use of the fact that 
the components of the RR potentials on the 
NS5-branes, which cannot be
gauged away are subject to a nonlinear equation similar to that of the
three form on the M5-brane.  To see that, start with an M5 with a large
transverse circle.  Then the three form is constrained by that
equation.  Making the circle small we can interpret the theory as type
IIA string theory and the three form is an RR potential, which is
subject to the same nonlinear equation.  By compactifying some of the
directions and using T-duality, a similar equation for the other RR
background fields is easily derived.  S-dualizing also leads to similar
relations on the worldvolume of the D5-brane (see Sec 6.4 ahead).

Consider the ODp limit, with the NS5-brane wrapped on a circle in the 
$p$ direction, with identification
$x^p\sim x^p+2\pi {R}_p$. 
Under T-duality in the $p$ direction, 
bulk quantities transform in the usual manner. 
Note, in particular, that the background RR fields change rank 
under the T-duality transformation 
\eqn\tdualbackrr{C_{01\ldots(p-1)}=2\pi {R}_p C_{01\ldots
p},~~~~C_{p\ldots 5}= { {R}_p \over 2\pi \aet} C_{(p+1)\ldots
5},}
and it is easily checked that the resulting scaling limit is that 
of the ODq (q=p-1) theory with unchanged effective scale $\aet$,  
on a circle of coordinate identification 
$x^p \sim x^p + 2 \pi \tilde{R}_p$ 
and dimensionless parameter $G_{o(p-1)}$ given by 
\eqn\tdual{\tilde{R}_p={\aet\over {R_p}}, ~~~~~G_{o(p-1)}^2
={\sqrt{\aet}\over {R}_p  }G_{o(p)}^2.}
Thus, like little string theories,  ODp theories inherit the
action of T-duality from the underlying string theory.

When an ODp theory is 
compactified on a torus whose metric is not diagonal (in the coordinate
system in which the the RR fields are as given in Table 3) 
the action of T-duality generates
RR fields of different ranks.  Decompactifying the T-dual torus 
one obtains the other six dimensional theories with several different RR
fields which have referred to  above.

\subsec{p=0}

The OD0 limit contains NS5-branes in the presence of a 
near critical 1-form gauge field $C_0= {1\over \epsilon
G_{o(0)}^2\sqrt{\aet}}(1-{\epsilon\over 2})$, leading to light
D0-branes.  In contrast to ODp theories with $p>0$ (and to 
NCOS and OM theories), the light excitations of 
the OD0 theory carry a conserved charge. Thus the OD0 may be 
studied in any of an infinite number of super-selection sectors, 
labelled by D0-brane charge. 

Since we are in IIA we can lift the NS5-brane to a
an M5-brane on a transverse circle with radius and Planck mass
\eqn\epsilono{R_{11}=g_{str}^{(0)}\sqrt{\tilde{\apm}}=\epsilon
G_{o(0)}^2\sqrt{\aet}\equiv \epsilon R, 
~~~~~M_p^3={1\over \epsilon^{3\over 2}G_{o(0)}^2\aet^{3\over 2}}
\equiv{1\over \epsilon^{3\over 2}}\met^3 .}

Choosing coordinates in the $11^{th}$ direction 
such that ($x^{11}\sim x^{11}+2 \pi R $), the 11 dimensional metric is 
\eqn\metm{ds_M^2=-(dx^0)^2+R_{11}^2({dx^{11} \over R}-C_0dx^0)^2 +\epsilon
dx_{\perp}^2 = \epsilon^2(dx^{11})^2 -\epsilon (dx^0)^2 -\epsilon
dx^{11}dx^0 +\epsilon dx_{\perp}^2.} 
Rescaling the unit of length by a factor of $\sqrt{\epsilon}$ 
(so that all lengths are larger, and all masses smaller, 
 by a factor of $\sqrt{\ep}$)
the metric, in the limit
$\epsilon\r 0$, takes the simple form  
\eqn\metmr{ds_M^2= -(dx^0)^2 - dx^{11}dx^0 +dx_{\perp}^2 .}
Note that the compactified direction $x^{11}$ is light-like. 
The bulk Planck scale in the new units is equal to $\met$.

In summary, the OD0 theory with N units of Do-brane charge is a DLCQ
compactification of M-theory,  (with 
Planck scale $\met={1\over G_{o(0)}^{2\over
3}\aet^{1\over 2}}$) with N units of DLCQ momentum, 
 in the presence of a transverse M5-brane. 
The periodic light-like coordinate has an identification of radius  
$R=G_{o(0)}^2\sqrt{\aet}$. 

The ODp theories for p$  > 1$ (like NCOS theories and \omt theory) have
excitations that are open $p$ branes. The presence of the appropriate
near critical RR potential keeps these open branes light, in the decoupling 
limit, provided they are appropriately oriented (branes with the opposite
orientation decouple). The counterpart of this statement  in 
the OD0 theories, is the fact that that D0 number must be positive; 
i.e. the familiar statement that the discrete momentum around a circle  
must be positive in a DLCQ compactification.

\subsec{p=1}

As pointed out above, the OD1 theory (with parameters 
$G_{0(1)}, \aet$) is S-dual to the 5+1
dimensional NCOS theory (with parameters $G_o, \ae$).  
%S-duality
%takes the NCOS theory to the OD1 theory on an NS5-brane with
%
%\eqn\nspar{g_{str}^{(1)}={1\over g_{str}}={\epsilon^{1\over 2}\over
%G_o^2},  ~~~~~ \tilde{\apm}=\apm g_{str}= \epsilon^{1\over 2}G_o^2\ae,
%~~~~g_{\m\n}=\eta_{\m\n},~~~ g_{MN}=\epsilon\delta_{MN}.}
%
The relationship between NCOS parameters (defined by Table 2) and ODp
parameters (defined by Table 3) is
\eqn\efdt{G_{o(1)}^2\equiv{1\over G_o^2},
~~~~~\aet\equiv \ae G_o^2.}
Notice the formal analogy to the transformation of the closed 
string quantities $\apm$ and $g_{str}$ under S-duality.

%Again, from S-duality, the expression for the RR two form field is 
%\eqn\rrtwo{\ep^{01}C_{01}={1\over 2\pi\ae\epsilon}(1-{\epsilon\over 2}).}
Note, of course, that D1-branes are exactly tensionless when 
$C_{01}$ takes its critical value as 
$\ep^{01}C^{crit}_{01}={1\over 2\pi\epsilon
G_{o(1)}^2\aet}={1\over 2\pi\ae\epsilon}$.
Unsurprisingly, the effective tension of D1-branes in the OD1 limit
is identical to that of NCOS strings in the S-dual 5+1 dimensional 
NCOS theory
\eqn\efftenodb{T_{eff}^{(1)}={1\over 4\pi\ae}\equiv {1\over 4\pi
G_{o(1)}^2\aet} .}

At low energies the OD1 theory reduces to a (5+1) dimensional gauge
theory with Yang-Mills coupling $g_{YM}^2=(2 \pi)^3\ae G_o^2=(2 \pi)^3
\aet$. Instantons in this gauge theory are strings (identified with
fundamental strings) in the low energy limit of the OD1 theory; these
strings consequently have tension $ \sim {1\over \aet}$.  
This yields an interpretation for the parameter $\aet$; it
sets the tension for closed little strings in OD1 theories.  As the
tension of a little string is unchanged under T-duality, this 
statement is true of all the ODp theories.

As an aside, consider the 5+1 dimensional NCOS theory, in the limit 
$\ae \r 0$, $G_o \r \infty$, $g^2_{YM}$ held fixed. 
$\T^{01}=\ep^{01}2 \pi \ae \r 0$ in this limit, and so (at least naively) 
the NCOS theory recovers 5+1 dimensional Lorentz invariance 
in this limit. Open string oscillator states are infinitely massive, 
and decouple in this limit. Thus we are left with an (apparently) 
Lorentz invariant 6 dimensional theory, whose low energy limit 
is Yang Mills theory. It is natural to conjecture
that the 5+1 dimensional NCOS theory reduces to the little string 
theory, in this limit. 

In terms of 
the variables of the dual OD1 theory, the limit of the previous 
paragraph is $G_{o(1)} \r 0$, $\aet$
held fixed. The conjecture above is thus equivalent to the 
assertion that the OD1 theory reduces to the IIB little string theory 
as $G_{o(1)}$ is taken to zero at fixed $\aet$.

It is also tempting to conjecture that the OD1 theory with 
scale $\aet$ and parameter $G_{o(1)}$ may be identified with the 
5+1 dimensional NCOS theory with scale $\ae={\aet \over G_{o(1)}^2}$ and 
coupling $G_o=G_{o(1)}$. Since the OD1
theory is S-dual to the $5+1$ NCOS, this conjecture amounts to
a self-duality conjecture for the $5+1$ dimensional NCOS theory.
The two theories have the same symmetries and reduce to
Yang-Mills theory at low energies. We have no
further evidence for this conjecture.

\subsec{p=2}

Since we are in IIA theory now, we can lift the OD2 limit to
M-theory. The scaling limit then involves an M5-brane on a point on
the transverse M-theory circle, in the presence of a near critical
$C_{012}$ potential. The Planck length, invariant length of the
$11^{th}$ circle and the C field are given, in terms of OD2
parameters, by
\eqn\mpar{R_{11}= \epsilon^{1\over 2}G_{o(2)}^2 \sqrt{ \aet},
~~~M_p^3={1\over  \epsilon G_{o(2)}^2 \aet^{{3 \over 2}}},
~~~\ep^{012}C_{012}={1\over 4\pi^2}M_p^3(1-{\epsilon\over 2}).}

Comparing with Table 1, we find that the OD2  theory is identical to
OM theory with $\me^3 ={ 1 \over 2 G_{o(2)}^2 \aet^{{3 \over 2}}}$, on
a transverse circle, of coordinate length $R=G_{o(2)}^2 \sqrt{ \aet}$.

Notice that the tension of a fundamental string in this theory is
given by ${1 \over 2 \pi}\me^3R={1 \over 4 \pi \aet}$, confirming the
interpretation of $\aet$ as a scale that sets the tension of
fundamental strings in ODp theories.

As a consistency check on some of the dualities described in this paper, 
we will relate the 5+1 dimensional NCOS theory 
(with parameters $\ae$ and $G_0$), compactified on a
circle of coordinate radius $R$, with a theory on a circle or 
radius ${1 \over R}$ through two different duality chains.

\item{a.} By T duality. As described in section 3, this leads to the 
4+1 dimensional NCOS theory with scale $\ae$, 
coupling $G_o^{'2}={G_o^2 \sqrt{\ae} \over R}$ and 
radius $\ae \over R$. 
\item{b.} By performing an S-duality, to the OD1 theory, 
with parameters $\aet=\ae G_o^2$, $G_{o(1)}^2={1 \over G^2_o}$ and 
coordinate radius $R$. Then performing a T-duality, to the OD2 theory 
with parameters $\aet=\ae G_o^2$,  $G_{o(2)}^2={\sqrt{\ae} \over 
G_o R}$
on a circle of radius $\ae G_o^2 \over R$. As argued above, this is 
\omt theory, with  $\me^3={R \over 2 G_o^2 \ae^2}$, with a transverse 
circle of coordinate length ${\ae \over R}$, compactified, 
in an `electric' direction, on a circle of 
length $\ae G_o^2 \over R$. However, using the formulae of section 4, 
\omt theory with these parameters on an 
electric circle is identical to the 4+1 dimensional
NCOS theory, with effective scale ${1 \over \ae}$ and coupling 
constant $G_o^{'2}=
G_o^2{\sqrt{\ae}\over R_2}$, on a circle of transverse size 
${\ae \over R}$, in agreement with the result of the simple T-duality
described in a) above.

\subsec{p=3}

Since here we are in the IIB theory, the OD3 theory  may be analyzed by
performing S-duality.  From Table 3 we see that the string coupling
$g_{str}^{(3)}$ and the RR 2-form $C_{45}$ are both of order one.
After an S-duality transformation we find a D5-brane with
$B_{45}={1 \over 2 \pi G_{o(3)}^2 \aet} $ of order one, 
$\tilde{g}_{str}^{(3)}={1\over
g_{str}^{(3)}}={1 \over G_{o(3)}^2}$ is of order one and  $\apm
=\tilde{\apm}g_{str}^{(3)}=\epsilon^{1\over 2}\aet G^2_{o(3)}$.
The metric is scaled as in Table 3. This is precisely the zero slope limit of 
\sw, leading to a low energy effective NCYM (with spatial noncommutativity
of rank 2). The noncommutativity parameter $\t$ and Yang Mills coupling 
$g^2_{YM}$ are given in terms of OD3 parameters by  
\eqn\lep{\t=2 \pi G_{o(3)}^2 \aet , ~~~~g^2_{YM}= (2\pi)^3 
\aet.}

Note that, as for $p=1$, the limit $G_{o(3)} \r 0$, $\aet$ fixed, 
takes the noncommutativity parameter $\t$ to zero 
(naively at least restoring Lorentz invariance) at fixed Yang Mills
coupling. Once again, it is natural to conjecture that this limit 
leads to the IIB little string theory. 

Of course, noncommutative Yang-Mills in 5+1 dimensions is
nonrenormalizable, and so quantum mechanically ill defined. 
The OD3 theory provides a completion of 6 dimensional NCYM.

\subsec{p=4}

In M-theory the NS-5brane is an M5-brane at a point on the 
$11^{th}$ circle. The light 4-branes of the OD4 theory
are also M5-branes- these M5-branes 
are wrapped on the eleventh
circle intersecting the transverse M5-brane in the directions $1\ldots
4$.  

The M theory parameters that correspond to the OD4 limit are
\eqn\mfive{R_{11}=G_{o(4)}^2\sqrt{\aet}, ~~~~M_p^3={1\over
\epsilon^{1\over 2} G_{o(4)}^2\aet^{3\over 2}}.}  
Note that $R_{11}$ is of order one.  
It is easy to check, directly in M-theory, that 
wrapped 5-branes of the appropriate orientation are light. $C_{01234}= {1\over
(2\pi)^4\epsilon G_{o(4)}^2{\aet}^{5\over 2}}(1-{\epsilon\over 2})$
lifts to $C_{01234,11}={1\over 2\pi R_{11}}C_{01234}= {1\over
(2\pi)^5}M_p^6(1-{\epsilon\over 2})$. This implies that the 5-branes
wrapped on the ${01234,11}$ directions are light with an effective
tension $T_{eff}^{(5)}={1\over 2(2\pi)^5}M_{eff}^6={1\over
(2\pi)^5}{\epsilon M_p^6\over 2}$ of order one.  (The dual field $C_5$
is of order one and hence only affects the geometry of the the $5-11$
plane by introducing a relative tilt in the coordinates).

\subsec{p=5}

Here we are in the IIB theory.  The theory has the full $SO(5,1)$ six
dimensional Lorentz invariance with $g_{\mu\nu}=\eta_{\mu\nu}$.  As we
see from Table 3, the string coupling diverges like
$\epsilon^{-{1\over 2}}$.  We are therefore tempted to use S-duality.
This converts the NS5-branes to D5-branes with a slope of order one.
Superficially, this is the scaling of the little string theory \lst.
But since the zero form $C$ is of order one, the string coupling after
the S-duality transformation does not go to zero, but 
continues to diverges like
$\epsilon^{-{1\over 2}}$.  So we end up with D5-branes with $C$ and
$\alpha'$ of order one but with a divergent string coupling.  

We conclude that this theory is not weakly coupled either before or
after the S-duality transformation.  This situation has already been
encountered in \lst, where the description of D5-brane excitations of
the little string theory was shown to be difficult.

\newsec{NCOS Theories at Strong Coupling in Various Dimensions}

The NCOS theories constructed in \refs{\gmms,\sst} are open string theories 
without a closed string sector. String perturbation theory  provides 
an effective description of these theories only at  weak open string
coupling $G_o$. It is natural to ask if the NCOS theories have 
a dual description that is weakly coupled at large $G_o$. 

Indeed, it has been argued in \gmms\ that the strongly coupled 3+1 
dimensional NCOS theory is dual to a weakly coupled spatially noncommutative 
theory. Further, in section 3 of this paper we have
argued that the strongly coupled 
4+1 dimensional  NCOS theory is well described by 6 dimensional \omt theory.
In this section we will examine the strong coupling behavior of the 
NCOS theories in 6 or lower dimensions. We will also present a  
test of the conjectured  dual description \gmms\  of the 3+1 dimensional 
NCOS theory.  
Our conjectures are summarized in 
Table 4 below.

\medskip
\cl{Table 4}
\bigskip
\centerline{\vbox{\offinterlineskip
\hrule
\halign{&\vrule#&
        \strut\quad#\quad\cr
height3pt&\omit&&\omit&\cr & Dimension &&
NCOS theory at Strong Coupling  
\hfil & \cr height3pt&\omit&&\omit&\cr \noalign{\hrule}
height3pt&\omit&&\omit&\cr & 1+1&& U(n) theory with single unit
of electric flux\hfil &\cr
& 2+1 && SO(8) invariant M2-brane theory\hfil &\cr
& 3+1 && Spatially noncommutative Yang Mills Theory \hfil & \cr
& 4+1 && \omt theory \hfil & \cr
& 5+1 && Self Dual \hfil & \cr
height3pt&\omit&&\omit&\cr}
\hrule} }
\medskip

\subsec{d=1+1}

Consider an infinite D-string in IIB theory with background metric, 
and closed string coupling as in Table 2.\foot{This case was discussed
in \igor\ as well as \gkp. The results of this section were developed 
partly in discussions 
with I. Klebanov, L. Susskind and N. Toumbas.}  The allowed values of the  
electric field on the D-string are quantized, and are given by 
(see for instance Eq 2.4 in \gmms) 
\eqn\allelfd{ {\ 2 \pi \apm \ep^{01} F_{01}
\over \sqrt{1+(2 \pi \apm)^2 F^2 }}=g_{str}n.}
\allelfd\ may be rewritten in terms of the quantities defined in 
section 3.1 as
\eqn\manip{1-{ \ep \over 2}={ng_{str} \over \sqrt{1+(ng_{str})^2}}.}
In the limit $ng_{str} \gg 1$ (the near critical limit) 
\eqn\manipp{ { \ep} ={1 \over (n g_{str})^2}
~~~{\rm i.e} ~~~\ae=\apm (n g_{str})^2.}
{}From table 2 we find 
\eqn\stc{G_o^2={1 \over n}.}
Thus we take the NCOS limit 
\eqn\ncost{ g_{str} \r \infty\, ~~~ \apm={\ae \over n^2g_{str} ^2}, 
~~~~~ n, \ae ~~{\rm fixed}}
to obtain the 1+1 dimensional NCOS theory with coupling constant 
$G^2_0={1 \over n}$ and string tension $\ae$. Notice that 
$G_o$ takes discrete values, and is 
bounded from above. Thus, unlike its higher dimensional counterparts, the 
1+1 dimensional NCOS theory is characterized by a discrete
(rather than continuous) parameter $n$, apart from a scale. Further, the 
1+1 dimensional NCOS theory cannot be taken to strong coupling. 

In order to obtain a dual description of this NCOS theory, 
we S dualize the background described above. We find a theory of 
$n$ D-strings with a single unit of electric flux, in a spacetime 
with 
\eqn\background{g'_{\m\n}={ \eta_{\m\n}\over g_{str}},
~~~~g'_{str}= {1 \over g_{str}}.}  
An electric field of the form $ F_{01}{\rm I}$ (I is the identity matrix)
on the D-strings is governed by the 
Born Infeld action for a U(1) field, times an extra factor of $n$ 
from the overall trace. In this picture, the value of $\d$ corresponding
to a single unit of flux may thus be determined from an equation 
analogous to \manip, 
\eqn\act{ 1-{ \ep \over 2}={{g'_{str} \over n } \over \sqrt{1+({g'_{str} \over n})^2}} 
={1 \over \sqrt{1+(n g_{str})^2}}.}
In the limit of interest $n g_{str} \r \infty$, so $ { \ep \over 2} \r 1$ and the 
background electric field is very far from criticality. Consequently, 
the open string coupling and metric are equal to the corresponding 
closed string quantities and  $\ae = \apm$. As $\apm \r 0$,
both open string oscillators and gravity decouple, and the 
D-string world volume theory is 1+1 dimensional U($n$) Yang Mills, 
with a single unit of electric flux,  governed by the action
\eqn\opo{S= { \pi \apm \over 2g'_{str}} 
\int d^2 x \sqrt{-g'}{ \rm  Tr}\left(
g^{'\m\a}g^{'\n\b}F_{\m\n}F_{\a\b} \right) 
={ \pi g_{str}^2 \apm \over 2} \int d^2 x 
{ \rm Tr} \left(\eta^{\m\a}\eta^{\n\b}F_{\m\n}F_{\a\b} \right).}
{}From \ncost, the Yang Mills coupling is
\eqn\ymc{ g^2_{YM}= {1 \over 2 \pi g_{str}^2 \apm}={n^2 \over 2\pi \ae}}
and remains fixed in the scaling limit. Note that this duality predicts 
that the 1+1 dimensional 
NCOS theory at the strongest allowed value of 
open string coupling $G_o^2=1$ is dual to a 
U(1) gauge theory, and so is secretly a free theory.

In summary,  
1+1 U($n$) Yang Mills with a single unit of electric flux, 
and gauge coupling $g^2_{YM}$ has a dual description as a
weakly coupled 1+1 dimensional NCOS theory with open string coupling
$G^2_0={1 \over n}$ and effective scale $\ae={n^2 \over  2\pi  g^{2}_{YM}}$!
In the rest of this section we will use this duality to 
study 1+1 dimensional large $n$ U($n$) gauge theory 
(with a single unit of flux)  at various energies.

As shown in 
\rwit, the 1+1 dimensional U($n$) theory with a single unit 
of electric flux reduces to a free U(1) theory at
low energies, as the SU($n$) part of this theory is massive.
The dual weakly coupled NCOS theory also reduces to a free U(1) theory 
at low energies; indeed it may be used to predict 
that the mass gap of the SU($n$) part of the U($n$) theory is 
${ 1\over \sqrt{\ae}}=\sqrt{2 \pi g^2_{YM} \over n^2}$.  

For a range of energies above the mass gap, and at large $n$,  
the duality derived in this section predicts that the 
the weakly coupled degrees of freedom of the 
gauge theory are open strings, with a tension 
\eqn\ymtension{{1 \over 4 \pi \ae}={g^2_{YM} \over 2n^2}}
and effective coupling $G_o^2={1 \over n}$. 
Indeed these stringlike excitations are easily identified in the 
gauge theory.  First recall why the $SU(n)$ part of the theory 
is gapped.
Excitations of the $SU(n)$ theory involve  
excitations of the scalars $X^I$ i.e. configurations involving  
separated D1-branes, as the gauge field has no dynamics.
However, because of the background electric field, all such 
excitations cost energy. For instance, 
a configuration in which a single D-string is taken to infinity
(the $SU(n)$ is Higgsed to $SU(n-1)\times U(1)$), has energy per 
unit length above that of the vacuum given by the BPS formula\foot{
A simple way to check the factors in this formula is to recall that the 
tension of a $(p, 1)$ string is 
${1 \over 2 \pi}\sqrt{{p^2 \over g_{str}^2}+1}$ and that in 1+1 
dimensions $g^2_{YM}={g_{str} \over 2 \pi \apm}$.}
\eqn\enpul{{g^2_{YM} \over 2 n} -{g^2_{YM} \over 2(n-1)}
\approx {g_{YM}^2\over 2 n^2} ~~~(n \gg 1)} 
as, in this limit, the electric flux is shared by 
$n-1$ rather than $n$ D-strings. But this tension 
agrees exactly with that of the NCOS string.
Thus an NCOS open string of length $L$ 
is identified with a configurations of the $n$ coincident D-branes  
of the gauge theory, in which the background electric flux
is shared between $n-1$ of the D-strings over a length $L$. 
The  last remaining D-string is free 
to fluctuate in the $R^8$ transverse to the D-branes over this 
segment of length $L$, but is bound to the branes everywhere else, 
resulting in an open string of length $L$ with Dirichlet boundary 
conditions. 

As with all string theories, the effective coupling of the 
NCOS theory grows with energy, and at energies much higher 
than the mass gap, the NCOS strings, (and, therefore, the `flux'
strings of the gauge theory, described in the previous paragraph)
 are strongly coupled. 
Indeed, at squared energies much larger than $g^2_{YM} n$, 
the usual W-bosons of the gauge theory constitute the weakly coupled 
variables for the gauge theory.

\subsec{d=2+1}

IIA theory in the NCOS limit of Table 2 may equivalently be described as  
M-theory on a circle of proper radius
$R_{11}={g_{str} \sqrt{\apm}}=G_o^2\sqrt{{\ae }}$ and a Planck mass
$\mp$ that goes to infinity $\mp^{-1}=g_{str}^{{1 \over 3}}  \sqrt{\apm}=
G_o^{{2 \over 3}} \apm^{{1 \over 3}}\ae^{{1\over 6}}.$
Recall that a  D2-brane in IIA theory, with no $F$ flux on its 
worldvolume,  maps to an M2-brane at a point on the $11^{th}$
circle. The dynamics of the gauge field on such a 2-brane maps to the 
dynamics of the compact scalar (representing the position of the M2-brane
in the $11^{th}$ direction) on the M2-brane world volume.
 
Now consider a D2-brane with a large electric field, as in Table 2.
One may choose to regard $F_{\m\n}$
instead of $A_{\m}$ as the dynamical variable in the
Born Infeld action on the D2-brane, if one simultaneously introduces  
a Lagrange Multiplier field  $\ph$ that  enforces the constraint $dF=0$;  
\eqn\bid{S=
{1\over (2 \pi)^2 g_{str}\apm^{{3 \over 2}}}
\int d^3x \sqrt{-\det\left(g_{MN}+ 2\pi \apm F_{MN}\right)}
+\half \int d^3 x  \sqrt{-g} \ep^{MNP}\p_{M} \ph F_{NP}.}
The equation of motion that results from varying this action with respect
to $F_{MN}$, specialized to the case of a diagonal metric $g_{MN}$ 
and a constant background electric field $F_{01}$, is 
\eqn\mtbd{\p_2 \ph ={1\over 4\pi^2 g_{str}\apm^{{3 \over 2}} }
{(2 \pi \apm)^2 \ep^{01}_{2} F_{01}
\over \sqrt{1+g^{11}g^{00}(2 \pi \apm F_{01})^2}}.}
The scalar field $\ph$ in \mtbd\ is dimensionless, and is compact of 
unit periodicity\foot{
The periodicity of $\ph$ may be deduced as follows. Consider the theory 
on  a circle in the 2 direction. The  RHS of \mtbd, integrated over the 
circle, is an integer, by the 3 dimensional analogue of \allelfd.
Hence $\oint dx^2 \p_2 \ph =n$.}  
 $\ph \equiv \ph +1$.
In the NCOS limit of Table 2, \mtbd\ may be rewritten as 
\eqn\mtbda{\p_2 \ph={1\over 2 \pi g_{str}\sqrt{\apm}}=
{1 \over 2 \pi R_{11}}.}

$2\pi R_{11}\phi=X^{11}$ is identified geometrically with the 
position of the brane in the $11^{th}$ direction. \mtbda\ implies that 
in the presence of a near critical electric field, the M2-brane 
tilts at an angle of 45 degrees in  the 2-11 coordinate plane 
\eqn\tilt{X^{11}=x^2.}
However, in the NCOS limit, $g_{22} \r 0$, so that, when angles are
measured in terms of physical distances, the M2-brane is oriented 
almost entirely in the 0,1,11 directions. More precisely, 
the 2 dimensional NCOS scaling limit
has a dual description in terms of an M2-brane  extended 
in the 0,1, directions, and spiraling around the 
2-11 cylinder\foot{This picture was also 
developed by O. Aharony and C. Vafa (private communication).}.
Successive turns of the spiral are separated by physical distance 
\eqn\separationp{\D X \approx 2\pi \sqrt{g_{22}} R_{11} 
=2\pi G_o^2\sqrt{\apm}.} 
The field $\psi$ on the worldvolume of the M2-brane, normalized so that
its kinetic term is $\half \int d^3x (\p \psi)^2$, is related to the physical 
displacement $X$ by $\psi = (2 \pi)^{-1}\mp^{{3 \over 2}} X$. 
Combining this with \separationp\ 
we find that the spacing between successive turns of the spiral in 
$\psi$ space is given by
\eqn\separationpsi{\D \psi = {G_o \over \ae^{{1 \over 4}}}}
and is finite in the NCOS limit. 

Interactions between successive windings may be ignored only for energies 
$\o \ll {G_o^2 \over \sqrt{\ae}}$. For $G_o \ll 1$ we thus obtain 
a complicated interacting theory at energy scale $\sim {1 \over \sqrt{\ae}}$.
On the other hand, in the strong coupling limit $G_o\r \infty$, 
interactions may be ignored at all energies, and the NCOS theory 
reduces to the free $SO(8)$ invariant theory of a 
single M2-brane.

\fig{When lifted to M theory, a D2-brane in the 0,1,2 directions with 
a near critical $F_{01}$ field is an 
M2-brane that that fills the 0-1 plane and 
spirals round the 2-11 cylinder. The pitch of the spiral is related 
to the open string coupling $G_o$ and goes to infinity at 
strong coupling.}{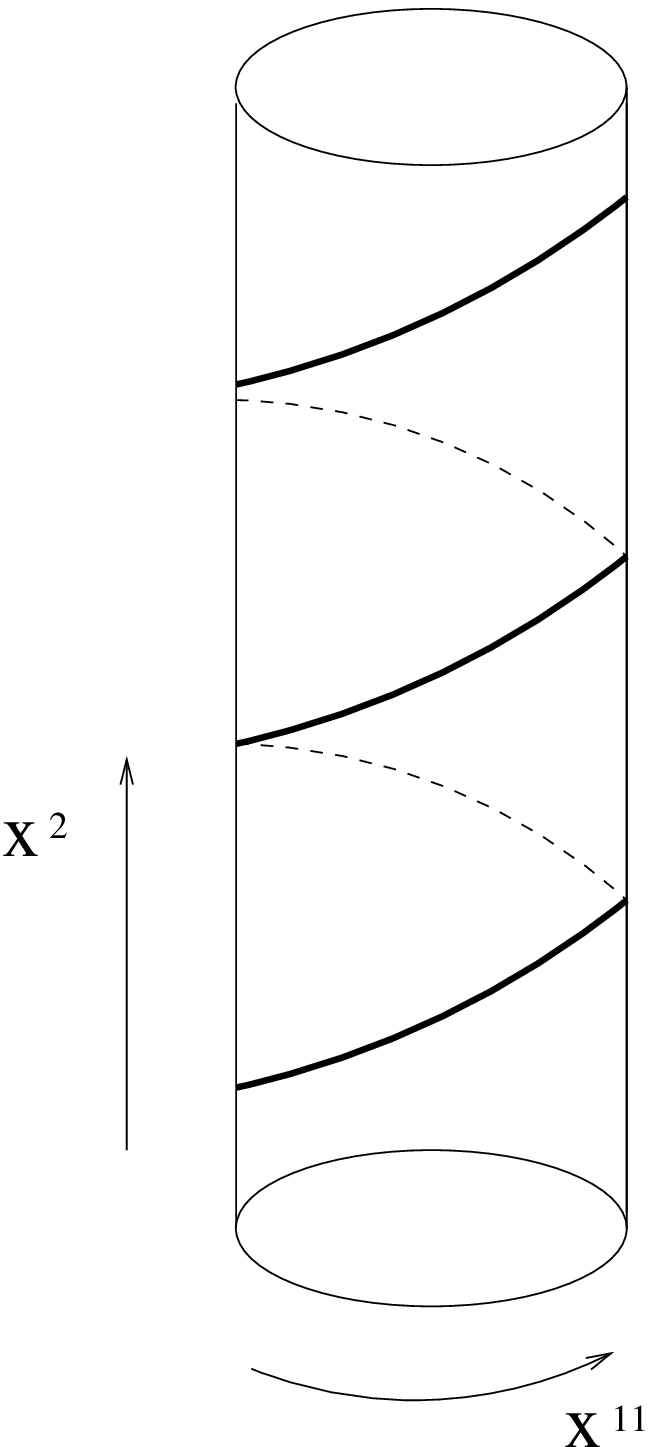}{1.5truein}

\newsec{Moduli Counting}

%\nref\wessez{wessez} 
%nref\sw{sw} 
\lref\persch{M.~Perry and J.H.~Schwarz, ``Interacting chiral gauge
fields in six-dimensions and Born-Infeld theory,'' Nucl. Phys. {\bf
B489} (1997) 47, hep-th/9611065.}

Since this paper involves branes in ${\cal R}^n$ in constant
background fields, we would like to make a few comments about such
backgrounds. We start by considering D$p$-branes in ${\cal
R}^{10}$. In the absence of the branes a constant NS $B$ field can
be gauged away. When the branes are present such a gauge
transformation changes the value of the field strength $F$ at
infinity $F(\infty)$, since only $F+B$ is gauge invariant.  We can
either fix the boundary conditions $F(\infty)$, and then $B$ is
meaningful, or gauge transform $B$ to zero and focus on the
background $F(\infty)$, or more generally discuss the gauge
invariant background $F(\infty)+B$.  The situation with M5-branes
in ${\cal R}^{11}$ is somewhat more interesting.  As for
D-branes, only ${\cal H}(\infty)= H(\infty)+C$ is gauge
invariant.  However, here there is a new element because the
field strength $\cal H$ is a constrained field.  For a single
M5-brane and a small and slowly varying $\cal H$ it has to be
selfdual, and for generic values of $\cal H$, which are still
slowly varying the condition is more complicated \wessez. This
equation should also be imposed on the boundary values ${\cal H}
(\infty)$ which characterize the problem \sw. This means that the
problem of M5-branes in flat ${\cal R}^{11}$ depends on $(6\times
5\times 4)/(2\times 3!)=10$ parameters, where the factor of $2$
arises from this generalized selfduality condition.

We now follow these parameters as the M5-branes are compactified
on a circle to become D4-branes in type IIA string theory. For
simplicity we consider a single M5-brane.  The six dimensional
three form $H$ field leads to a two form field strength $F=\oint H$
in five dimensions and a three form.  The M5-brane equation
relates them and determines one of them in terms of the other
\persch.  Hence, we can take the degrees of freedom to be
ordinary gauge fields with field strength $F$.

For slowly varying fields the dynamics of the D4-brane is well
known to be controlled by the Lagrangian
\eqn\ldbi{L=h(F+B) +C\wedge (F+B),}
where $h$ is the DBI Lagrangian (for a review see \ref\tse{A.~A.~Tseytlin,
``Born-Infeld action, supersymmetry and string theory,''
hep-th/9908105.}).  The term
proportional to the RR field $C$ can be dropped when $C$ is a
constant since it is a total derivative. \ldbi\ is invariant
under the electric gauge symmetries
\eqn\electricg{\eqalign{
&\delta A=d \lambda_e +\Lambda_e\cr &\delta B = - \Lambda_e.}} We
note that for constant $B$ and $C$ we are free to specify 20
independent parameters as well as the boundary conditions
$F(\infty)$.  However, only $F(\infty)+B$ is gauge invariant and
meaningful, and the terms proportional to $C$ do not affect the
local dynamics.  Therefore the problem is characterized only by
10 parameters, exactly as for its ancestor M5-brane.

Let us perform a duality transformation on the Lagrangian \ldbi.
We do that by viewing $F$ as an independent field and by
introducing a Lagrange multiplier $V$ to implement the Bianchi
identity for $F$.  The Lagrangian $L$ is replaced by
\eqn\ldbipp{h(F+B) + C\wedge (F+B) -V\wedge dF.}
Next, we integrate by parts to replace \ldbipp\ by
\eqn\ldbip{h(F+B)+(C+dV)\wedge(F+B).}
The Lagrangian \ldbip\ is invariant under the magnetic gauge
symmetries
\eqn\magneticg{\eqalign{
&\delta V=d \lambda_m +\Lambda_m\cr
&\delta C = - \Lambda_m.}}
The equation of motion of $F$ is algebraic
\eqn\feom{h'(F+B)+C+F_D=0; \qquad \qquad F_D=dV.}
It has a number of consequences:
\item{1.} The two field strengths $F$ and $F_D$ are the two form
and three form which are obtained by dimensional reduction of the
M5-brane field $H$, and $B$ and $C$ are the dimensional reduction
of the higher dimensional $C$ field.  Here we see explicitly how
they are related.  As shown in \persch, equation \feom\ is the
dimensional reduction of the generalized selfduality equation
of \wessez.
\item{2.} Equation \feom\ can be solved $F+B=f(C+F_D)$ and $F$
can be integrated out to
express the the theory in terms of the dual variables as
\eqn\ldbipd{L_{dual}=h(f(C+F_D)) + (C+F_D)\wedge f(C+F_D).}
We see that the dual Lagrangian is independent of $B$
and that the dependence on the background RR field $C$ is
nonlinear.
\item{3.} In \ldbi\ the constants $C$ are arbitrary, as they
multiply total derivative terms.  However, they do affect the
boundary conditions of the dual variables $V$.  To see that,
we should examine \feom\ at infinity
\eqn\feomi{h'(F(\infty)+B)+C+F_D(\infty)=0.}
Clearly, the value of $F_D(\infty)$ depends on $C$.  As for the
electric variables, only $C+F_D(\infty)$ is gauge invariant and
physical.  Furthermore, these constants are determined in terms
of the electric constants $F(\infty)+B$.  Therefore, the problem
is characterized by $(5\times 4)/2=10$ parameters, exactly as for
the M5-brane we started with.

It is straightforward to repeat this analysis for D3-branes in ${\cal
R}^{10}$. Here the background depends on $(4\times 3)/2=6$ parameters
from the NS $B$ field, which can be interpreted as boundary values of
the field strength $F$ at infinity. The $(4\times 3)/2=6$ parameters
in the RR $C$ field multiply total derivative terms and do not affect
the dynamics.  Hence, the total number of independent gauge invariant
two form parameters is six. S-duality transformation can be performed
as above except that now $F_D$ and $C$ are two forms.  Again, equation
\feomi\ relates the boundary values of $F$ and $F_D$ in terms of the
parameters $B$ and $C$.  

A natural gauge choice is $B=C=0$, and then the boundary conditions on
the dynamical fields are $h'(F(\infty))+{}^*F_D(\infty)=0$.  With this
choice S-duality exchanges nonzero background electric field with
background magnetic field.  Another natural choice is such that the
dynamical fields vanish at infinity $F(\infty) =F_D(\infty)=0$.  Then
the NS and the RR fields are related by $h'(B)+{}^*C=0$.  Here
S-duality exchanges the related values of $B$ and $C$.  The local
dynamics depends only on the value of the NS $B$ field, so that
S-duality can be described as mapping $B \to -{}^*h'(B)$ (it is
straightforward to check that this transformation is a $Z_2$
transformation).

\centerline{\bf Acknowledgements}
  We are grateful to O. Aharony, 
V. Balasubramanian, T. Banks, J. De Boer, I.Klebanov,  
B. Pioline, A. Sen, S. Sinha, L. Susskind, N. Toumbas, C. Vafa 
and especially J. Maldacena for useful discussions.  
We would like to thank Aki Hashimoto for pointing out an 
error in an earlier version of this paper. 
The character \omt was generated using the freely downloadable 
software package
Devanagari for TeX (devnag).
NS thanks the Racah Institute of
the Hebrew University for hospitality when some of this work was done.
The work of R.G. ,  S.M. and A.S 
was supported in part by DOE grant DE-FG02-91ER40654.
The work of N.S. was supported in part by DOE grant \#DE-FG02-90ER40542.

\listrefs
\end

%% file: dnmacs.tex
%
%    dnmacs.tex v1.5
%
%    Plain TeX macros for Devanagari for TeX package
%    Copyright (C) 1991-1998  University of Groningen, The Netherlands
%
%    Author:   Frans J. Velthuis <velthuis@rc.rug.nl>
%
%    This program is free software; you can redistribute it and/or modify
%    it under the terms of the GNU General Public License as published by
%    the Free Software Foundation; either version 1, or (at your option)
%    any later version.
%
%    This program is distributed in the hope that it will be useful,
%    but WITHOUT ANY WARRANTY; without even the implied warranty of
%    MERCHANTABILITY or FITNESS FOR A PARTICULAR PURPOSE.  See the
%    GNU General Public License for more details.
%
%    You should have received a copy of the GNU General Public License
%    along with this program; if not, write to the Free Software
%    Foundation, Inc., 675 Mass Ave, Cambridge, MA 02139, USA.
%

\font\smalldn=dvng8
\font\ninedn=dvng9
\font\dvng=dvng10
\font\halfdn=dvng10 scaled\magstephalf
\font\bigdn=dvng10 scaled\magstep1
\font\largedn=dvng10 scaled\magstep2
\font\hugedn=dvng10 scaled\magstep3
\def\sethyph#1{
\hyphenchar\smalldn=#1
\hyphenchar\ninedn=#1
\hyphenchar\dvng=#1
\hyphenchar\halfdn=#1
\hyphenchar\bigdn=#1
\hyphenchar\largedn=#1
\hyphenchar\hugedn=#1}
\font\smallcr=cmr8
\font\ninecr=cmr9
\font\halfcr=cmr10 scaled\magstephalf
\font\bigcr=cmr10 scaled\magstep1
\font\largecr=cmr10 scaled\magstep2
\font\hugecr=cmr10 scaled\magstep3
\let\rsize=\rm
\newcount\chnum
\newdimen\itdim
\newdimen\dnblskip
\newif\ifdnmode
\def\subscr#1{\/\itdim=\lastkern
\unkern\kern-\itdim \lower\dp0 \hbox to\itdim{#1\hfil}}
\def\dnsmall{\let\pdn=\smalldn\let\rsize=\smallcr%
\dnblskip=12pt\ifdnmode\dn\fi}
\def\dnnine{\let\pdn=\ninedn\let\rsize=\ninecr%
\dnblskip=13pt\ifdnmode\dn\fi}
\def\dnnormal{\let\pdn=\dvng\let\rsize=\rm%
\dnblskip=15pt\ifdnmode\dn\fi}
\def\dnhalf{\let\pdn=\halfdn\let\rsize=\halfcr%
\dnblskip=16pt\ifdnmode\dn\fi}
\def\dnbig{\let\pdn=\bigdn\let\rsize=\bigcr%
\dnblskip=18pt\ifdnmode\dn\fi}
\def\dnlarge{\let\pdn=\largedn\let\rsize=\largecr%
\dnblskip=22pt\ifdnmode\dn\fi}
\def\dnhuge{\let\pdn=\hugedn\let\rsize=\hugecr%
\dnblskip=26pt\ifdnmode\dn\fi}
\def\dn{\dnmodetrue\pdn\baselineskip=\dnblskip
\chnum=0
\loop\catcode\chnum=11
\ifnum\chnum<12\advance\chnum by1
\repeat
\chnum=14
\loop\catcode\chnum=11
\ifnum\chnum<31\advance\chnum by1
\repeat
\catcode127=11
\tolerance=10000
\pretolerance=10000}
\def\0{\llap{\char13}}
\def\1{\llap{\char32}}
\def\2{\llap{\char92}}
\def\3#1w{{\char"#1}}
\def\4{\llap{\char123}}
\def\5{\llap{\char125}}
\def\6#1{\setbox0=\hbox{#1}#1\subscr{\char126}}
\def\7#1{\setbox0=\hbox{#1}#1\subscr{\char0}}
\def\8#1{\setbox0=\hbox{#1}#1\subscr{\char1}}
\def\9#1{\setbox0=\hbox{#1}#1\subscr{\char2}}
\def\qa#1#2{\setbox0=\hbox{#1}#1\subscr{\char253\kern1.5ex\lower1.25ex
\hbox{\char#2}\kern-1.5ex}}

\def\qva{\kern0.5ex\2\kern-0.5ex}
\def\qvb{\kern1ex\0\kern-1ex}
\def\qvc{\kern1ex\rdt\kern-1ex}
\def\?{\llap{\char3}}
\def\<{\llap{\char4}}

\def\rdt{\llap{\char19}}
\def\dnnum{\let\nstyle=d}
\def\cmnum{\let\nstyle=r}
\cmnum
\def\rn#1{\if\nstyle r{\rsize #1}\else#1\fi}

\sethyph{-1}
\let\pdn=\dvng
\dnblskip=15pt